\magnification=1200
\baselineskip 8truemm
\hsize 15truecm
\def\parn{\par\noindent}
\def\eps{{\cal E}}
\def\Rt{r_{\rm t}}
\def\Ro{R_{\circ}}
\def\Io{I_{+}}

$\;$
\vskip 3truecm
\centerline{\bf INVERSION OF THE ABEL EQUATION}
\medskip
\centerline{\bf FOR TOROIDAL DENSITY DISTRIBUTIONS} 
\bigskip\bigskip
\centerline{L. Ciotti}
\centerline{Osservatorio Astronomico di Bologna}
\centerline{via Ranzani 1, 40127 Bologna (Italy)}
\centerline{e-mail: ciotti@astbo3.bo.astro.it}
\vfill (with 1 figure)\eject
\centerline{\bf ABSTRACT}
\bigskip\parn
In this paper I present three new results of astronomical interest
concerning the theory of Abel inversion. 1) I show that in the case
of a spatial emissivity that is constant on toroidal surfaces and
projected along the symmetry axis perpendicular to the torus'
equatorial plane, it is possible to invert the projection
integral. From the surface (i.e. projected) brightness profile one
then formally recovers the original spatial distribution as a function
of the toroidal radius. 2) By applying the above--described inversion
formula, I show that if the projected profile is described by a
truncated off-center gaussian, the functional form of the related
spatial emissivity is very simple and -- most important -- nowhere
negative for any value of the gaussian parameters, a property which is
not guaranteed -- in general -- by Abel inversion. 3) Finally, I show how
a generic multimodal centrally symmetric brightness distribution can
be deprojected using a sum of truncated off-center gaussians,
recovering the spatial emissivity as a sum of nowhere negative
toroidal distributions.
\bigskip\parn
{\it keywords}: Methods: analytical -- 
                Methods: numerical -- 
                Methods: data analysis
\vfill\eject
\centerline{\bf 1. INTRODUCTION}
\bigskip\parn
A common problem in astronomy is the deprojection of a given surface
brightness distribution. Unfortunately this problem is degenerate,
i.e., different spatial emissivities can originate the same surface
(projected) brightness distribution. As a consequence, any inversion
procedure is invariably based on a (more or less) arbitrary choice of
the underlying geometry of the spatial emissivity. A help to this
unpleasant situation comes from two guide rules: in choosing the
geometry of the spatial emissivity one uses symmetry properties (if
any) of the brightness distribution, and after deprojection discards
the assumed geometry if this produces a somewhere negative spatial
emissivity.

One of the simplest cases is given by a surface brightness
characterized by central symmetry, i.e., described by a function
$I(R)$ of the projected radius $R$. The natural assumption is a
spherically symmetric spatial emissivity, $\eps=\eps(r)\geq 0$, where
$r$ is the spherical radius.  It is a well known result that in this
case the projection operator and its deprojection are given by an Abel
integral equation:
$$I(R) = 2 \int_R^{\Rt}{\eps (r) rdr \over \sqrt{r^2 - R^2}},
\eqno (1)$$
and 
$$\eps (r)=-{1\over \pi}\left
            [\int_r^{\Rt} {d I(R)\over dR} {dR\over\sqrt{R^2
            -r^2}}-{I(\Rt)\over\sqrt{\Rt^2-r^2}}\right ],
\eqno (2)$$  
where $\Rt$ is the truncation radius, i.e., $I(R)=0$ for $r >\Rt$, and
for untruncated distributions $\Rt\to\infty$. It is important to note
that if $I(\Rt)>0$ the recovered emissivity is (weakly) divergent at
the edge of the sphere, due to the second therm on the r.h.s. of
eq. (2). A fundamental problem of Abel inversion is that the
positivity of the recovered distribution is not guaranteed. As an
example of this fact, one can consider $I(R)$ to be a gaussian or an
off-center gaussian: while in the first case the deprojected
emissivity -- being still a gaussian -- is everywhere positive (see,
for applications, Bendinelli, Ciotti, \& Parmeggiani 1993, and
references therein), in the second case it can be easily shown that
$\eps(r)$ diverges negatively for $r\to 0$.

What other symmetries for $\eps$ are compatible with a surface
brightness $I=I(R)$ and are still of astrophysical interest? The
natural generalization of the spherical symmetry is the cylindrical
one. Many astronomical objects are characterized by this geometry:
lenticular and spiral galaxies, accretion disks, plasma tori,
planetary rings, some planetary nebula etc.. In this paper we are
interested in particular in the problem of deprojecting a $I(R)$ which
presents some off--center maxima, and so the natural assumption about
the emissivity distribution $\eps$ is the toroidal one.

In Section 2 I show that in the case of a toroidal emissivity
distribution projected along its symmetry axis it is possible to
generalize the inversion formula (2). Then I show that in the
particular case of a projected brightness described by a truncated
off--center gaussian, $\eps$ has an extremely simple functional form,
and at variance with the spherically symmetric case, {\it it is
nowhere negative for any choice of the gaussian parameters}. In
Section 3 I propose the use of the found physically admissible
$I$--$\eps$ pair to obtain a (finite) series expansion of the
emissivity for a {\it generic} centrally symmetric surface brightness
profile. In a following paper (Bendinelli et al. 1999), an application
of this method to the Planetary Nebula (PN) A~13 is described. This PN
is just one case among many others of astrophysical interest to which
the method produces useful results on the object's structure.
\bigskip
\centerline{\bf 2. THE INVERSION FORMULA FOR TOROIDAL SYMMETRY}
\bigskip\parn
Let us start by obtaining the analogous formula of eq. (1) for the
projection along the symmetry axis of an emissivity distribution
stratified on toroidal surfaces. The natural coordinates are the
cylindrical, $(R,\varphi,z)$, where $z=0$ is the torus equatorial
plane.  The relation between cylindrical and toroidal coordinates
$(r,\varphi,\vartheta)$ is given by
$$R=\Ro+r\sin\vartheta,\quad\varphi=\varphi,\quad z=r\cos\vartheta ,
\eqno (3)$$
(see Figure 1).  
\vskip 1 truecm
\parn
{\bf Figure 1} {\it The describing parameters of toroidal geometry, and their
relations used for the projection and deprojection.}
\bigskip\parn
We assume independence\footnote{$^*$}{The request of
independence of $\eps$ from $\varphi$ is unnecessary, and the
following discussion can be easily generalized to distributions
$I=I(R,\varphi)$, obtaining $\eps=\eps(r,\varphi)$.}  of $\eps$ from
$\vartheta$ and $\varphi$, and so each isoemissivity surface is
labeled by its toroidal radius $r=\sqrt{z^2+(R-\Ro)^2}$ around the
circle of radius $\Ro$ placed in the equatorial plane. Moreover, for
the sake of generality let us assume that the emissivity is non-zero
only for $0\leq r\leq\Rt\leq\Ro$. For $\Rt=\Ro$ we have a so-called
{\it full torus}. With the previous assumptions, $I(R)\geq 0$ for
$|R-\Ro|\leq\Rt$ and $I(R)=0$ outside. From very simple geometric
arguments one obtains the analogous formula of eq. (1):
$$I(R)=2\int_{|R-\Ro|}^{\Rt} {\eps(r) rdr\over\sqrt{r^2 -(R-\Ro)^2}}.
\eqno (4)$$
With respect to the variable $R$, the l.h.s. of eq. (4) is a
generalized Abel integral, but it is not directly invertible, since
$(R-\Ro)^2$ is not strictly increasing with $R$ (see Gorenflo \&
Vessella 1991, p.24). Since the function's profile is symmetric with
respect to $\Ro$, we can use a branch of $I$ (i.e., we use $R\geq
\Ro$) and we define a new variable $s:= R-\Ro\geq 0$ with
$\Io(s):= I(\Ro +s)$. In this way, the integral is formally
identical to the projection operator in eq. (1) and can be inverted:
$$\eps(r)=-{1\over\pi}\left [\int_r^{\Rt} {d\Io\over
ds} {ds\over\sqrt{s^2-r^2}}-{\Io(\Rt)\over\sqrt {\Rt^2 -r^2}}\right ].
\eqno (5)$$
Thus, we proved that {\it assuming the surface brightness distribution
to be the projection on its equatorial plane of a toroidally
stratified emissivity, it is formally possible to invert the
projection integral, recovering the emissivity as a function of the
toroidal radius}.
\bigskip
\centerline{\bf 3. TRUNCATED OFF--CENTER GAUSSIANS}
\bigskip\parn
As an application of eq. (5) we invert a surface brightness profile
described by an off--center gaussian truncated at $\Rt$:
$$I(R)=S\left\{\exp\left [-{(R-\Ro)^2\over 2\sigma^2}\right ]-
     \exp \left (-{\Rt^2\over 2\sigma^2}\right )\right\},
     \quad |R-\Ro|\leq\Rt .
\eqno (6)$$
Note that $\Io(\Rt)=0$, and so the unpleasant divergence at the edge
of the torus is avoided.  The total luminosity $L$
associated to $I$ is given by its surface integral over the annulus
$\Ro-\Rt\leq R\leq\Ro+\Rt$:
$$L=S\Ro\sigma\,(2\pi)^{3/2}\left [
  {\rm Erf}\left({\Rt\over\sqrt{2}\sigma}\right )-
  {\Rt\sqrt{2}\over\sigma\sqrt{\pi}}\exp\left(-{\Rt^2\over 2\sigma^2}\right )
                         \right ], 
\eqno (7)$$ 
where ${\rm Erf}(x)=2/\sqrt{\pi}\int_0^x\exp(-t^2)dt$ is the error
function.  The formula for the luminosity density given using
eqs. (5)-(6) results to be:
$$\eps (r)={S\over
\sqrt{2\pi}\sigma}\exp\left(-{r^2\over 2\sigma^2}\right )
         {\rm Erf}\left ({\sqrt{\Rt^2-r^2\over 2\sigma^2}}\right ).
\eqno (8)$$
The deprojection formula can be verified by inserting eq. (8) in
eq. (4) and then evaluating the integral. Note that the spatial
emissivity given by eq. (8) is {\it everywhere positive}, for any
choice of the gaussian parameters $(S,\Ro,\Rt,\sigma)$. One can then
conclude with the following analogy: {\it off--centered gaussians in
toroidal symmetry correspond to centered gaussians in spherical
symmetry}.

Having found a nowhere negative $I$-$\eps$ pair that satisfies the
Abel inversion for a particular toroidal density distributions, the
successive step is the natural extension of a previous work
(Bendinelli et al. 1993), where a multigaussian expansion of observed
profiles is described, under the assumption of spherical symmetry:
here it is sufficient to remember that by using gaussian functions one
avoids the computational difficulty of direct Abel numerical
inversion, which belongs to the class of unstable inverse problems
(Gorenflo \& Vessella 1991; Craig \& Brown 1986).

So one can assume that a given profile $I(R)$ is expanded as a sum of
truncated off-center gaussians with different parameters
$(S,\Ro,\Rt,\sigma)_i$ for $i=1,...,N$. The parameters are easily
computable using the Newton--Gauss regularized method which is a
powerful iterative non linear fitting technique (Bendinelli et al.
1987; Bendinelli 1991). Clearly the parameters are constrained to
satisfy the request that the integral over all the projection plane of
the fitted brightness distribution must be equal to the original one,
i.e. the total luminosity of the system $L=\sum_i L_i$, where $L_i$
are given by eq. (7). The associated spatial emissivity is then
approximated by the sum of $N$ distributions as in eq. (8).
The developed technique 
will be applied to the deprojection of the
galactic PN A~13, which is characterized by a well defined ring-shaped
morphology (Bendinelli et al. 1999).
\bigskip
\centerline{\bf 4. CONCLUSIONS}
\bigskip\parn
The results of this work are the following:
\parn
1) It is shown that it is possible to extend the classical Abel inversion
      for spherically symmetric systems to toroidal density distributions 
      projected along the torus symmetry axis. 
\parn
2) It is also shown that an off-center truncated gaussian function gives 
      rise to a well behaved spatial emissivity, i.e., the emissivity is a very
      simple function of the toroidal radius. More important, the emissivity 
      is non-negative over all the space for any choice of its parameters, a 
      property not guaranteed by the Abel inversion.
\parn
3) Finally, it is proposed the use of the found $I$-$\eps$ pair for the 
      recovering of the spatial emissivity of any centrally symmetric projected
      distribution as a sum of toroidal density distribution, after a 
      non-linear parameter fitting. 
\bigskip
\centerline{\bf 5. ACKNOWLEDGEMENTS}
\bigskip\parn
I would like to thank O. Bendinelli, G. Parmeggiani, and
L. Stanghellini for useful discussions.  This work was partially
supported by the contracts ASI-95-RS-152 and ASI-ARS-96-70.
\bigskip
\centerline{\bf 6. REFERENCES}
\bigskip\parn
Bendinelli, O., Ciotti, L., Parmeggiani, G., and Stanghellini, L., 1999,\par 
                in preparation.\parn 
Bendinelli, O., 1991, 
                ApJ, 366, 599.\parn
Bendinelli, O., Parmeggiani, G., Piccioni, A., and Zavatti, F., 1987, 
                AJ, 94, 1095.\parn
Bendinelli, O., Ciotti, L., and Parmeggiani, G., 1993, 
                A\&A, 279, 668.\parn
Craig, I.J.D., Brown, J.C., 1986, 
                ``Inverse Problems in Astronomy'', Adam Hilger,\par 
                Bristol.\parn
Gorenflo, R., Vessella, S., 1991, 
                ``Abel Integral Equations'', Springer--Verlag,\par
                Berlin.\parn
\vfill\eject
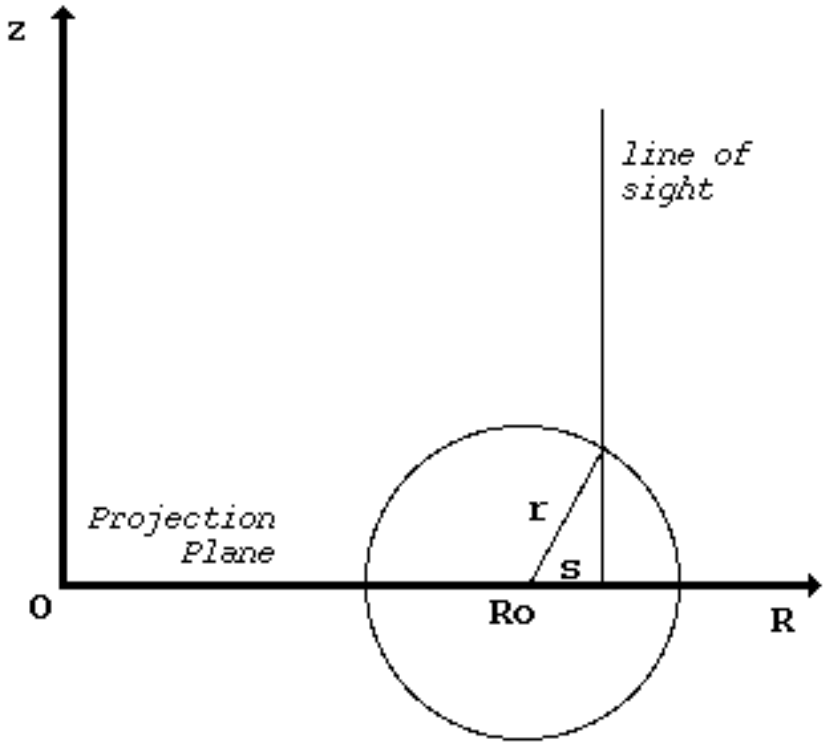
\bye